\documentclass[12pt,a4paper]{article}
\usepackage{amssymb,amsmath,amscd,epsfig}
\title{Purity sieve for models with  factorizable interactions}

\author{
 Oleg Lychkovskiy
\hspace*{2mm}$^{\rm a,b}$
\\ ${\rm ^a}$ {\small\it Institute for Theoretical and Experimental Physics}\\
{\small\it 117218, B.Cheremushkinskaya 25,
Moscow, Russia}\\
${\rm ^b}$ {\small\it Moscow Institute of Physics and Technology
}\\{\small\it 141700, 9, Institutskii per., Dolgoprudny, Moscow
Region, Russia}}

\date{}

\begin{document}
\maketitle

\newcommand{\be}{\begin{equation}}
\newcommand{\ee}{\end{equation}}

\newcommand{\K}{\mathbf{k}}
\newcommand{\n}{\mathbf{n}}
\newcommand{\q}{\mathbf{q}}
\newcommand{\e}{\mathbf{e}}
\newcommand{\A}{\mathcal{A}}

\newcommand{\p}{\tilde{p}}
\newcommand{\x}{\tilde{x}}

\newcommand{\la}{\langle}
\newcommand{\ra}{\rangle}

\newcommand{\cH}{\cal H}
\newcommand{\cS}{\cal S}
\newcommand{\cE}{\cal E}

\newcommand{\hPS}{\hat P^{\cal S}}
\newcommand{\hPE}{\hat P^{\cal E}}
\newcommand{\hH}{\hat H}
\newcommand{\hHE}{\hat H^{\cal E}}
\newcommand{\hHS}{\hat H^{\cal S}}
\newcommand{\hV}{\hat V}
\newcommand{\hS}{\hat S}
\newcommand{\hE}{\hat E}

\newcommand{\dens}{\hat \rho}
\newcommand{\densS}{\hat \rho^{\cal S}}
\newcommand{\densE}{\hat \rho^{\cal E}}
\newcommand{\meandens}{\hat{ \bar \rho}}
\newcommand{\meandensS}{\hat{ \bar \rho}^{\cal S}}

\newcommand{\tr}{\mathrm{tr}}
\newcommand{\trS}{\mathrm{tr}_{\cal S}}
\newcommand{\trE}{\mathrm{tr}_{\cal E}}
\newcommand{\Pu}{\mathfrak P}

\newcommand{\im}{\mathrm{i}}

\begin{abstract}
Evolution of purity in case of factorizable interaction between an open system and an environment is investigated. We derive a simple expression for purity decrease at the first instants of evolution (when purity is close to unity). We find that purity at very small times is maximal when an initial state of an open system coincides with one of the eigenstates of the interaction operator, no matter how weak the interaction is. On the other hand it is widely known that in general eigenstates of the interaction are not pointer states. Therefore the procedure of selecting pointer states by purity maximization (known as "purity sieve") should not rely on short-time purity behavior. We propose a modification of the purity sieve criterion which approximately takes into account purity evolution at longer times. As an example of its applicability we recover known results for pointer states of a particle undergoing quantum Brownian motion; we point out that the criterion is not applicable for some other models, however. It is argued that the proposed modified purity sieve may be used for selecting pointer states of a particle undergoing decoherence through collisions.
\end{abstract}

\section{Introduction}
Purity sieve \cite{Zurek 1993}\cite{Zurek et al 1993} is one of the criteria for selecting pointer states of an open quantum system, i.e. the states most robust against decoherence. In this work purity sieve is investigated in the case of factorizable interaction between an open system and an environment. This case is relevant for the majority of models usually considered in the context of decoherence, for example for quantum Brownian motion model, spin-boson model and some spin-spin models. The paper is organized as follows. In the next section we briefly review the conventional setting and the basic concepts of the decoherence program, outline the role of purity and remind the definition of the purity sieve. In section 2 we analyze the short-time behavior of purity. Also a modified (with respect to the canonical purity sieve) criterion for selecting pointer states is proposed in this section. In section 3 we introduce a model which has the following features: (1) it may be reduced to the "almost factorizable" form; (2) under certain assumptions decoherence in this model may be regarded as decoherence through collisions; (3) the particular case of this model is quantum Brownian motion. We use the proposed modified criterion to find pointer states for this model.

\section{Some basics of decoherence. }
In this section we give a very brief and incomplete sketch of the basic setting of the decoherence program. One may find a number of extensive reviews (e.g. \cite{Zurek}) and pedagogical introductions to the topic (e.g. \cite{Schlosshauer book}). In particular, a profound list of references to the original works is given in \cite{Zurek}.

Consider a System (with Hilbert space $\cS$), which interacts with an Environment (with Hilbert space $\cE$). The total Hamiltonian of the combined closed system (with Hilbert space $\cH=\cS \otimes \cE$)  reads
\be
\hH=\hHS+\hHE+\hV,
\ee
$\hHS,~\hHE$ and $\hV$ being self-Hamiltonians and interaction Hamiltonian correspondingly. The usage of superscripts $\cS$ and $\cE$ here and in what follows is self-explanatory.

The state of the combined system $\cH$ may be described by a state vector $\Psi$ or by a density matrix $\dens.$ The states of the System $\cS$ and the Environment $\cE$ are described by the reduced density matrices
\be\label{densS}
\densS \equiv \tr_{\cE} \dens,~~~\densE \equiv \tr_{\cS} \dens.
\ee
Density matrices evolve according to the master equations:

\be\label{Liuville unitary}
\frac{d}{dt}\dens=-\im [\hH,\dens],
\ee
\be\label{Liuville nonunitary}
\frac{d}{dt}\densS=-\im [\hHS,\densS]-\im~\tr_{\cE}[\hV,\dens].
\ee

According to the Schmidt theorem, at any time $\densS$ and $\densE$ may be diagonalized in some basis:
\be\label{decoherence}
\densS (t)=\sum_i~ p_i(t) \hPS_i (t),
\ee
\be\label{decoherence environment}
\densE (t)=\sum_i~ p_i(t) \hPE_i (t).
\ee
Here $\{\hPS_i (t)\}$ and $\{\hPE_i (t)\}$ are systems of mutually orthogonal projectors, and $\{p_i(t)\}$ is a set of non-negative numbers (probabilities).

An essential ingredient of the decoherence program is the following \\
{\bf \emph{Decoherence Hypothesis:}} for generic System, sufficiently large Environment, generic interaction and (almost) any initial conditions,  the projectors $\{\hPS_i(t)\}$ almost always belong to some set of fixed projectors, which correspond to the so-called {\it pointer states}. \\

In other words, the reduced density matrix is diagonal almost always in the basis of pointer states. Exceptional periods of time when the diagonalization basis substantially deviates from the pointer basis are short and rare.  If the initial state of the System is far from pointer, the first moments of the evolution form such an exception: the System {\it decohere} in the mixture of the pointer states during this period.

One of the main problems of the decoherence theory is to find pointer states, given a Hamiltonian $\hH$ of the combined system $\cH.$ One of the helpful tools for this is {\it purity sieve} \cite{Zurek 1993},\cite{Zurek et al 1993}. In order to introduce purity let us first note that there is a simple method to obtain the probability $p_{\rm max},$ which is by definition the largest one among all $p_i,$ and the corresponding projector $\hPS_{\rm max},$ given the reduced density matrix $\densS$:

\be\label{first pointer state}
\hPS_{\rm max}= \lim_{k \rightarrow \infty}\left[ (\densS)^k/\trS(\densS)^k \right],
\ee
\be\label{maximal probability}
p_{\rm max}= \trS \hPS_{\rm max} \densS = \lim_{k \rightarrow \infty}\left[ \trS(\densS)^{k+1}/\trS(\densS)^k \right].
\ee
In practice, approximate expressions for $\hPS_{\rm max}$ and $p_{\rm max}$ may be used:

\be\label{first pointer state aprox}
\hPS_{\rm max}\simeq  (\densS)^k/\trS(\densS)^k
\ee
\be\label{maximal probability aprox}
p_{\rm max}\simeq   \trS(\densS)^{k+1}/\trS(\densS)^k.
\ee
Here the larger is $k,$ the better is approximation.

Assume now that at $t=0$ the state of the System is pure, $\densS(0)=\mid \psi^{\cS}\ra\la \psi^{\cS}\mid.$ Then $p_{\rm max}$  initially equals 1, and stays close to 1 during some time. In this case even $k=1$ appears to be a good approximation, and one gets

\be\label{maximal probability very aprox}
p_{\rm max}\simeq   \trS(\densS)^2.
\ee
The quantity in the r.h.s. of the above equation is called {\it purity}. We denote it by $\Pu:$
\be
\Pu\equiv\trS(\densS)^2.
\ee
Sometimes $-\Pu$ is called "linear entropy".

For the pure state $\densS=\mid \psi^{\cS}\ra\la \psi^{\cS}\mid$ purity equals 1, and it decreases in the course of evolution.
The rate of the decrease differs for different initial states. It is natural to assume (in the spirit of the {\it Decoherence Hypothesis}) that for pointer states this rate should be small. The small rate of purity loss implies that the corresponding state is stable against decoherence. This reasoning suggests that in order to find pointer states one should minimize the purity decrease rate with respect to the initial pure states. This procedure is called {\it purity sieve}.

To summarize, purity sieve is a procedure for selecting pointer states of the System. It comes to the following two steps:\\
1. calculating purity $\Pu$ as a functional of the initial state of the System $\psi^{\cS},$ a functional of the initial state of the Environment $\densE,$ and a function of time $t;$
\footnote{This implies that the combined closed system starts its evolution from the initial product state $\dens=|\psi^{\cS}\rangle\langle\psi^{\cS}|\otimes\densE.$}
\\
2. maximizing $\Pu$ with respect to $\psi^{\cS}$ for some fixed (reasonable) $t$ and $\densE.$\\
The states which provide the local maxima should be the pointer states. The procedure makes sense if the result weakly depends on $\densE$ and $t,$ provided $t$ is greater than some characteristic time, called {\it decoherence time}.

\section{Purity sieve for factorizable interactions}

A number of models, investigated in the context of decoherence studies, have factorizable interaction Hamiltonian (see e.g. \cite{Zurek et al 1993}, \cite{Zurek 1982}-\cite{Cuccietti}):
\be
\hV=\hS\hE.
\ee
Such form of interaction allows to simplify the master equation (\ref{Liuville nonunitary}):
\be\label{ME fact}
\frac{d}{dt}\densS=-\im [\hHS,\densS]-\im~[\hS,\tr_{\cE}(\hE\dens)].
\ee
Using this expression, we calculate the first two derivatives of purity at $t=0:$

\be
\dot \Pu = 0,
\ee
\be\label{Pu''}
\ddot \Pu = -4\delta \hE^2 \delta \hS^2.
\ee
Here the dispersions squared are calculated for the initial states of the System and the Environment:
\be
\delta \hE^2\equiv \trE ~\hE^2\densE-(\trE~\hE\densE)^2,
\ee
\be\delta \hS^2\equiv \la \psi^{\cS}\mid \hS^2 \mid \psi^{\cS}\ra - \la \psi^{\cS}\mid \hS \mid \psi^{\cS}\ra^2.
\ee

The first derivative of purity vanishes at $t=0$ merely because the evolution starts from the state with maximal purity, $\Pu=1.$

The formula for the second derivative, (\ref{Pu''}), is a result of the straightforward calculation; it appears to be less lengthy if one uses the interaction representation instead of the Schroedinger representation.

The short-time behavior of purity is quadratical in time and reads
\be\label{quadratical decrease}
\Pu\simeq 1 - 2\delta \hE^2 \delta \hS^2 t^2.
\ee

A noticeable fact is that in the case of factorizable interactions the second derivative at $t=0$ also factorizes. As a consequence, as long as eq.(\ref{quadratical decrease}) holds, purity decrease is minimal for eigenstates of the interaction operator $\hS$, which give $\delta \hS^2=0$ and $\Pu=1-O(t^3).$ Does this mean that purity sieve selects the eigenstates of $\hS$ to be the pointer states of the System? In general, the answer is negative. For example, if the interaction is weak compared to the self-Hamiltonian of the System, pointer states are known to be the eigenstates of the self-Hamiltonian $\hHS$ \cite{Paz and Zurek}.

\begin{figure}
\centerline{\epsfig{file=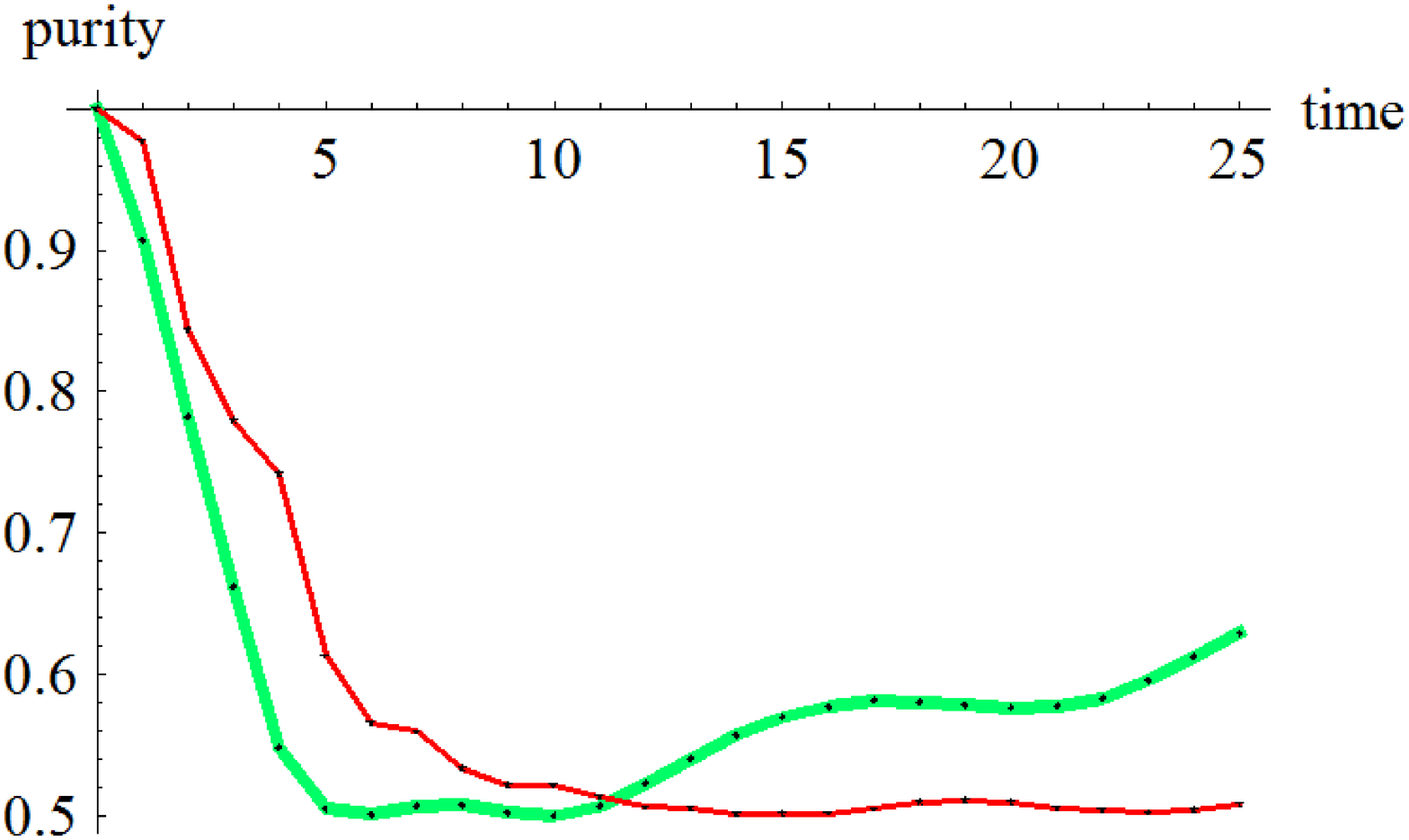, width=10.5cm, height=5cm}}
\caption{Purity as a function of time for the central spin model (\ref{1}) and for two different initial states, $|0z\rangle$ (thick green line) and $|0x\rangle$ (thin red line). The parameters of the model are $N=6,~\omega=1,~\epsilon=0.1.$ The initial state of the environment is a mixture of all possible states with equal weights ($\densE=1/2^N$).}
\label{}
\end{figure}

To illustrate this situation let us consider a simple central-spin model as an example:
\be\label{1}
\hH = \frac{\omega}{2}\sigma_z+\sum_{i=1}^N \frac{\omega}{2}\sigma_z^i + \epsilon\sigma_x\sum_{i=1}^N \sigma_x^i.
\ee
Here $\sigma_x$ and $\sigma_z$ correspond to a central spin, $\sigma_x^i$ and $\sigma_z^i$ correspond to $N$ bath spins, constants $\omega$ and $\epsilon$ characterize the strengths of self-Hamiltonian and interaction correspondingly. $\Pu(t)$ is calculated numerically for two initial directions of the central spin, along $0z$ axis, $|0z\rangle$ (the eigenstate of the self-Hamiltonian $\hHS=\frac{\omega}{2}\sigma_z$), and along $0x$ axis, $|0x\rangle$ (the eigenstate of the interaction operator $\hS=\sigma_x$). The result for $N=6,~\omega=1,~\epsilon=0.1$ is shown in Fig.1. One can see that at first moments the eigenstate of the interaction operator, $|0x\rangle,$ indeed "wins" the competition, providing minimal purity decrease. However, at longer times the situation is not so clear. Purity evolution is no longer described by the quadratic law (\ref{quadratical decrease}); purity sieve does not unambiguously select any state. We conclude that eq.(\ref{quadratical decrease}) correctly describes short-time behavior of the considered model, but purity sieve is hardly applicable in this case.

Another example is a particle which undergoes decoherence through collisions with other particles. The interaction hamiltonian in this case is a function of distance between particles. According to eq.(\ref{quadratical decrease}) sharply localized states (coordinate delta-functions) maximize  purity at small times. However, such wave functions are known to instantly spread over large distances. This implies an abrupt decrease of purity at slightly longer times. Therefore, in this case one also can not use eq.(\ref{quadratical decrease}) to select pointer states.

The examples considered show that maximizing short-time expression (\ref{quadratical decrease}) does not, in general, provide correct pointer states. However, the later example gives a hint how one can try to upgrade a selection criterion. The key quantity in eq.(\ref{quadratical decrease}) is the dispersion squared of the interaction operator, $\delta \hS^2.$ In order to maximize purity we minimized $\delta \hS^2,$ calculated for the initial state $\psi^{\cS}.$
\footnote{The factor $\delta \hE^2$ is expected to take some typical value for large environments; it should not affect the result.}
A straightforward generalization is to try to calculate the dispersion of the interaction operator for longer times, to average it and then -- to minimize it. Thus, a modified criterion for selecting pointer states may be proposed, which comes to the minimization of
\be\label{Pu approx}
\int_0^t dt' \delta \hS^2 (t'),
\ee
 where
 \be
\delta \hS^2\equiv \la \psi^{\cS}(t)\mid \hS^2 \mid \psi^{\cS}(t)\ra - \la \psi^{\cS}(t)\mid \hS \mid \psi^{\cS}(t)\ra^2,
\ee
and $\psi^{\cS}(t)$ is an evolving state of the System (which is assumed to be approximately pure). It satisfies a Shroedinger-like equation
\be\label{ME approx}
i\frac{d}{dt}\psi^{\cS}(t)= (\hHS+\kappa\hS)\psi^{\cS}(t), ~~~\kappa\equiv \trE \hE\densE.
\ee

This criterion appears to work satisfactory in the case when the interaction operator is a coordinate operator, ${\cS} =\hat{x}.$ This may be verified by considering specific examples (see the next section). On the other hand, one can easily check that it does not work, for example, in spin-spin model (\ref{1}). The reason for such diverse behavior seems to be connected with the singularity or non-singularity of the interaction operator. However, this question is far from being clearly understood. It will be discussed elsewhere.

\section{Purity sieve for quantum Brownian motion and for decoherence through collisions}

In this section we give an example of implementation of the the proposed modified selection criterion.

The following Hamiltonian is considered:

\be\label{our model Hamiltonian}
\hH=\frac{p^2}{2M}+\frac{1}{2}M\Omega^2x^2+\sum_{k=1}^N(\frac{p_k^2}{2m}+\frac{1}{2}m\omega^2x_k^2)
+\sum_{k=1}^N U_1(x_k-x)+ \sum_{k>j}^N  U_2(x_k-x_j)
\ee

It describes one particle of mass $M$ and $N$ particles with masses equal to $m$ in the oscillator potential wells.  All particles interact with each other through potentials $U_1$ and $U_2.$  If this potentials are collision potentials, the model describes decoherence through collisions. If the potentials are quadratic, this is a quantum Brownian motion (QBM) model.

Let us rewrite this Hamiltonian in an "almost factorizable" form. For this we use
the following transformation of variables (\ref{our model Hamiltonian}):

\begin{tabular}{l}
$\x=x$\\
$\x_k=x_k-x$\\
$\p=p+\sum p_k$\\
$\p_k=p_k.$
\end{tabular}\\
In new variables the Hamiltonian reads

$$\hH=\hHS+\hHE+\hV,$$
\be
\hHS=\frac{\p^2}{2M}+\frac{1}{2}(M\Omega^2+Nm\omega^2) \x^2
\ee
\be
\hHE=\sum_{k=1}^N(\frac{\p_k^2}{2m}+\frac{(\sum\p_k)^2}{2M}+\frac{1}{2}m\omega^2\x_k^2)
+\sum_{k=1}^N U_1(\x_k)+ \sum_{k>j}^N  U_2(\x_k-\x_j)
\ee
\be\label{interaction in our model}
\hV=-\frac{\p\sum\p_k}{M}+m\omega^2\x \sum \x_k.
\ee


One can see that the interaction Hamiltonian is a sum of two factorizable terms now.  We could make some further assumptions in order to make one of the terms in (\ref{interaction in our model}) negligible and get rid of it. However, it appears to be unnecessary. Indeed, let us calculate the integral (\ref{Pu approx}), which is subject to minimization, separately for both terms:
$$\int_0^T dt' \delta \p^2 (t')=1/2(M^2\tilde\Omega^2\delta x(0)^2+\delta p(0)^2)T,$$
$$\int_0^T dt' \delta \x^2 (t')=1/2(\delta x(0)^2+\delta p(0)^2/M^2\tilde\Omega^2)T,$$
where
\be
\tilde\Omega^2\equiv \Omega^2+\frac{Nm}{M}\omega^2,
\ee
and integration is performed over the oscillator period $T$. We see that due to $p-x$ symmetry in the oscillator Hamiltonian two contributions appear to coincide up to a constant factor. One can prove that the interference term vanishes. Minimizing the obtained expression one gets the pointer states to be those with
\be
\delta\x~\delta\p=1/2,
\ee
\be\label{coordinate uncertainty}
\delta \x^2=(2M\tilde\Omega)^{-1}.
\ee

Thus pointer states are minimum uncertainty states with the coordinate uncertainty defined by (\ref{coordinate uncertainty}).

Previously this result was obtained for the QBM \cite{Zurek et al 1993}. From the mathematical point of view calculations in \cite{Zurek et al 1993} coincide with those in the present paper, although a different approach, based on the approximate master equation, was used in \cite{Zurek et al 1993}.

Approaches, which are also very different from the presented one, are usually used to study  decoherence through collisions (see, e.g. \cite{Paz et al},\cite{Zeh}). These studies normally predict the pointer states to be phase-space localized states (as demanded by the quantum-to classical correspondence principle), which agrees with our result.

Although our result for the model (\ref{our model Hamiltonian}) seems to be in general reasonable, one should be cautious when applying it to specific forms of Hamiltonian with certain functions $U_1$ and $U_2.$ For example, if the interaction vanishes, $U_1=0,$  the very conception of pointer states makes no sense, but our selection criterion does not "feel" it and yet gives minimum-uncertainty states as pointer states. As it was mentioned above, the applicability conditions of the proposed criterion are not entirely explored.

\section*{Acknowledgements}
The author is grateful to organizers of the Fourth International Workshop DICE2008 for hospitality. The
work was financially supported by the Dynasty Foundation scholarship, RF
President grant NSh-4568.2008.2, RFBR grants 07-02-00830-a and RFBR-08-02-00494-a.

\end{document}